\documentstyle[preprint, aps,epsfig,psfig]{revtex}         
\newcommand{\pt}{\mbox{$^3P_2$}}

\begin{document}
\title{
\large\bf Properties of 
the Tensor Mesons $f_2(1270)$ and $f^\prime_2(1525)$}

\author{ De-Min Li${}^b$\footnote{ E-mail: 
lidm@hptc5.ihep.ac.cn/lidm@alpha02.ihep.ac.cn}, Hong Yu${}^{a,b,c}$, 
Qi-Xing Shen${}^{a,b,c}$\\
{\small \em  $^a{}$CCAST (World Lab), P.O.Box 8730, Beijing 100080, P.R. 
China\\ 
$^b{}$Institute of High Energy Physics, Chinese Academy of
Sciences, \\
P.O. Box 918 (4), Beijing 100039, P.R. China\footnote{Mailing 
address}\\ 
$^c{}$Institute of Theoretical Physics, Chinese Academy of 
Sciences, Beijing 100080, P.R. China}}
\maketitle
\vspace{0.5cm}

\begin{abstract}
In the $f_2(1270)-f^\prime_2(1525)$ mixing framework, the isoscalar 
singlet-octet mixing
angle for $1~\pt$ tensor nonet is determined to the value of $27.5^\circ$
and the decays of the two states are investigated. Comparing the
predicted results of the decays of the two states with the available
experimental data, we find that the predicted results are in good
agreement with the experimental data, which shows that the $f_2(1270)$ and 
$f^\prime_2(1525)$
wave functions don't need other components such as glueballs or the 
$2~\pt$, $3~\pt$, ... $q\overline{q}$.
 \end{abstract}

\vspace{0.5cm}

\newpage


\section*{1. Introduction}

In the 1200$\sim$1800 MeV mass range, one expects that a tensor 
glueball, the $1~\pt$, $2~\pt$, $3~\pt$ and $1~^3F_2$ nonets exist. At 
present, in this mass range, 13 isoscalar tensor states are claimed 
to exist experimentally\cite{PDG1}. 
The state $f_2(1430)$ was claimed to be found in the data on the 
double-Pomeron-exchange reaction $pp\rightarrow p_f(\pi^+\pi^-)p_s$ at 
$\sqrt{s}=63$ GeV in an experiment R807 at CERN ISR\cite{R807}, however, 
recent experiments on the same reaction do not see any evidence for 
$f_2(1430)$\cite{9903}. Its existence needs further experimental 
confirmation. Among the other states, $f_2(1270)$ and
$f^\prime_2(1525)$ are well known as the ground tensor states. Above 
$f^\prime_2(1525)$, none of the reported isoscalar tensor states can be 
definitely 
assigned to be the member of the  $2~\pt$, $3~\pt$, $1~^3F_2$ nonets or the 
tensor glueball\cite{PDG}. Recently, it is controversial
that whether $f_2(1270)$ and
$f^\prime_2(1525)$ need glueballs components or not. Ref.\cite{6282}
sifted these overpopulated isoscalar tensor states using Schwinger-type mass 
relations derived from a mass matrix in which only the 
$q\overline{q}$-glueball coupling was considered. Inputing the masses of 
some observed but possibly confused states, Ref.\cite{6282} found that 
the 
physical tensor mesons $f_2(1270)$ and $f^\prime_2(1525)$ have a substantial 
glueball 
content. Ref.\cite{5193} assumed that $f_2(1640)/f_2(1710)$ is the 
quarkonia-glueball mixing state, and investigated the mixing of $f_2(1270)$, 
$f^\prime_2(1525)$ and $f_2(1640)/f_2(1710)$ in the $1~\pt$ 
$N=(u\overline{u}+d\overline{d})/\sqrt{2}$, $1~\pt$ $S=s\overline{s}$ and 
$G=gg$ basis. Ref.\cite{5193} suggested that the absence of the gluonic 
components in the tensor mesons $f_2(1270)$ and $f^\prime_2(1525)$ due to 
the predicted 
branching ratios are incompatible with the experimental data. We favor 
the suggestion that $f_2(1270)$ and $f^\prime_2(1525)$ don't need other 
components 
such as glueballs. Since the mass of the lowest
lying tensor glueball predicted by lattice QCD is larger than
2 GeV\cite{LQCD}, which is far from the masses of $f_2(1270)$ and
$f^\prime_2(1525)$, one can
qualitatively expect that the mixing between the tensor glueball and the
$1~\pt$ $q\overline{q}$ would be rather little\cite{CHAO}. 
However, we propose that the states chosen in 
Ref.\cite{5193} are too arbitrary. First, the spin of $f_J(1710)$ has been 
controversial\cite{god}. Close {\sl et al.} argued that $f_J(1710)$ would be a 
$q\overline{q}$ state if $J=2$ and $f_J(1710)$ would be a mixed 
$q\overline{q}$ glueball having a large glueball component if 
$J=0$\cite{Close}. 
However, evidence for spin 0 have accumulated recently in all 
production modes for $f_J(1710)$\cite{9903,Spin}, and the state 
$f_J(1710)$ with $J=0$ has been cited by Particle Data Group 
2000 (PDG 2000)\cite{PDG}. Second, there is not any evidence that 
$f_2(1640)$ has 
advantages over other states to be assigned as a tensor glueball mixing 
with $1~\pt$ $q\overline{q}$. In the viewpoint of A.V. Anisovich {\sl et 
al.}\cite{AV}, it 
seems reasonable to assign $f_2(1640)$ as the first excitation of 
$f_2(1270)$ 
and $f_2(1810)$ as the first excitation of $f^\prime_2(1525)$. If so, it 
is obviously 
unreasonable to discuss the mixing of $f_2(1270)$, $f^\prime_2(1525)$ and 
$f_2(1640)$ in the 
$1~\pt$ $N$, $1~\pt$ $S$ and $G$ basis. Third, according to the masses of 
the states chosen in Ref.\cite{5193}, the mass of the lowest lying tensor 
glueball is determined to be about 1.5 GeV. Such a low mass tensor 
glueball would be very difficultly accommodated by lattice QCD which 
predicts the mass of the tensor glueball is larger than 2 
GeV\cite{LQCD}. Finally, as mentioned above, except for $f_2(1270)$ and 
$f^\prime_2(1525)$, 
none of the reported isoscalar tensor states can be definitely assigned 
to be the $2~\pt$, $3~\pt$, $1~^3F_2$ $q\overline{q}$ or the tensor 
glueball. There 
thus are not any convincing reasons to only choose $f_2(1640)/f_2(1710)$ but 
not other state to mix with $f_2(1270)$ and $f^\prime_2(1525)$. 

In this work, we shall avoid all the isoscalar tensor states which are 
confused or need further experimental confirmation, and adopt a simple 
model to quantitatively check that whether the $f_2(1270)$ and 
$f^\prime_2(1525)$ wave 
functions need other components such as glueballs or the $1~\pt$, $2~\pt$, 
$3~\pt$, ... $q\overline{q}$ or not.

\section*{2. Mixing model}
In the $1~\pt$ $|N\rangle=|u\overline{u}+d\overline{d}\rangle/\sqrt{2}$,
$1~\pt$ $|S\rangle=|s\overline{s}\rangle$ basis, the mass-squared matrix
describing the quarkonia-quarkonia mixing can be written as
follows\cite{Matrix}:
\begin{equation}
M^2=\left(\begin{array}{cc}
M^2_N+RA&\sqrt{R}A\\
\sqrt{R}A&M^2_S+A
\end{array}\right),
\end{equation}
where $M_N$ and $M_S$ are the masses of the bare states $1~\pt$ $|N\rangle$
and $1~\pt$ $|S\rangle$, respectively; $A$ is a mixing parameter which
describes the transition amplitude of $s\overline{s}$ annihilation and
reconstitution via intermediate gluons states. The appearance of $R$ 
means that we consider the possibility that the transition between 
$q\overline{q}$ and
$q^\prime\overline{q^\prime}$ is flavor-dependent. Here we assume that the 
physical states
$|f_2(1270)\rangle$ and $|f^\prime_2(1525)\rangle$ are the eigenvectors of 
the matrix $M^2$
with the eigenvalues of $M^2_{f_2(1270)}$ and $M^2_{f^\prime_2(1525)}$, 
respectively. Diagonalizing the mass matrix $M^2$, we have
\begin{equation}
UM^2U^\dagger=\left(\begin{array}{cc}
M^2_{f_2(1270)}&0\\
0&M^2_{f^\prime_2(1525)}
\end{array}\right),
\end{equation}
the physical states $|f_2(1270)\rangle$ and $|f^\prime_2(1525)\rangle$ 
can be written as \begin{equation}
\left(\begin{array}{c}
|f_2(1270)\rangle\\
|f^\prime_2(1525)\rangle
\end{array}\right)=
U\left(\begin{array}{c}
|N\rangle\\
|S\rangle
\end{array}\right),
\end{equation}
where the unitary matrix $U$ can be given by
\begin{equation}
\left(\begin{array}{cc}
X_{f_2(1270)}&Y_{f_2(1270)}\\
X_{f^\prime_2(1525)}&Y_{f^\prime_2(1525)}
\end{array}\right)=
\left(\begin{array}{cc}
\sqrt{R}A/C_1&(M^2_{f_2(1270)}-M^2_N-RA)/C_1\\
\sqrt{R}A/C_2&(M^2_{f^\prime_2(1525)}-M^2_N-RA)/C_1
\end{array}\right)
\end{equation}
with $C_1=\sqrt{RA^2+(M^2_{f_2(1270)}-M^2_N-RA)^2}$,
$C_2=-\sqrt{RA^2+(M^2_{f^\prime_2(1525)}-M^2_N-RA)^2}$.
It follows from Eqs. (1) and (2) that 
\begin{eqnarray}
&&M^2_N+M^2_S+RA+A=M^2_{f_2(1270)}+M^2_{f^\prime_2(1525)},\\
&&(M^2_N+RA)(M^2_S+A)-RA^2=M^2_{f_2(1270)}M^2_{f^\prime_2(1525)}.
\end{eqnarray}
From Eqs. (5) and (6), $A$ and $R$ can be derived as
\begin{eqnarray}
A=\frac{(M^2_{f_2(1270)}-M^2_S)(M^2_S-M^2_{f^\prime_2(1525)})}{M^2_S-M^2_N},\\
R=\frac{(M^2_{f_2(1270)}-M^2_N)(M^2_N-M^2_{f^\prime_2(1525)})}{M^2_{f_2(1270)}-M^2_S)(M^2_{f^\prime_2(1525)}-M^2_S)}.
\end{eqnarray}
Apart from $M_{f_2(1270)}=1.2754$ GeV and $M_{f^\prime_2(1525)}=1.525$ 
GeV\cite{PDG}, we take
$M_N=M_{a_2(1320)}=1.318$ GeV and $M_S=1.55$ GeV\cite{6282} as input, the 
numerical form of the unitary matrix can be given by
\begin{equation}
U=\left(\begin{array}{cc}
X_{f_2(1270)}&Y_{f_2(1270)}\\
X_{f^\prime_2(1525)}&Y_{f^\prime_2(1525)}
\end{array}\right)=
\left(\begin{array}{cc}
-0.991&-0.135\\
0.135&-0.991
\end{array}\right),
\end{equation}
then the physical states $|f_2(1270)\rangle$ and $|f^\prime_2(1525)\rangle$ 
can be given by 
\begin{eqnarray}
&&|f_2(1270)\rangle=-0.991|N\rangle-0.135|S\rangle,\\
&&|f^\prime_2(1525)\rangle=0.135|N\rangle-0.991|S\rangle.
\end{eqnarray}
If we re-express the two physical states in the Gell-Mann
$|8\rangle=|u\overline{u}+d\overline{d}-2s\overline{s}\rangle/\sqrt{6}$,
$|1\rangle=|u\overline{u}+d\overline{d}+s\overline{s}\rangle/\sqrt{3}$,
$|f_2(1270)\rangle$ and $|f^\prime_2(1525)\rangle$ can be read as
\begin{eqnarray}
&&|f^\prime_2(1525)\rangle=\cos\theta_T|8\rangle-\sin\theta_T|1\rangle,\\
&&|f_2(1270)\rangle=\sin\theta_T|8\rangle+\cos\theta_T|1\rangle,
\end{eqnarray}
with $\theta_T=27.5^\circ$, which is in good agreement with the value of
$28^\circ$ given by PDG 2000\cite{PDG}.

\section*{3. Decays}
 
For the hadronic decays of $f_2(1270)$ and $f^\prime_2(1525)$ into two 
pseudoscalar mesons,
we consider the coupling modes as indicated in Fig. I: i) the direct
coupling of the quarkonia components of the initial mesons to the final 
pseudoscalar mesons occurring as the leading order decay mechanism, and 
ii) the coupling of
the quarkonia of the initial mesons to the final pseudoscalar mesons via 
intermediate gluons states occurring as the next leading order decay
mechanism. Based on these coupling modes, the effective Hamiltonian
describing the hadronic decays of $f_2(1270)$ and $f^\prime_2(1525)$ into 
two pseudoscalar mesons can be described as\cite{Hamil}
\begin{eqnarray}
H_{eff}=g_1{\bf Tr}(f_FP_FP_F)+g_2{\bf Tr}(f_F){\bf Tr}(P_FP_F),
\end{eqnarray}
where $g_1$ and $g_2$ describe the effective coupling strengths of the
coupling modes i) and ii), respectively; $f_F$ and $P_F$ are $3\times3$
flavor matrixes describing the $q\overline{q}$ components of the initial
tensor mesons and
the final pseudoscalar mesons, respectively. Based on the mixing 
scheme mentioned in section 2, $f_F$ can written as 
\begin{equation}
f_F=\left(\begin{array}{ccc}
\sum\limits_{i}\frac{X_i}{\sqrt{2}}i&0&0\\
0&\sum\limits_{i}\frac{X_i}{\sqrt{2}}i&0\\
0&0&\sum\limits_{i}Y_ii
\end{array}\right)~~(i=f_2(1270),~f^\prime_2(1525)).
\end{equation}
$P_F$ can be written as
\begin{equation}
P_F=\left(\begin{array}{ccc}
\frac{\pi^0}{\sqrt{2}}+\alpha\eta+\beta\eta^\prime&\pi^+&K^+\\
\pi^-&-\frac{\pi^0}{\sqrt{2}}+\alpha\eta+\beta\eta^\prime&K^0\\
K^-&\overline{K^0}&-\sqrt{2}\beta\eta+\sqrt{2}\alpha\eta^\prime
\end{array}\right),
\end{equation}
where
\begin{eqnarray}
\alpha=(\cos\theta_p-\sqrt{2}\sin\theta_p)/\sqrt{6},~~
\beta=(\sin\theta_p+\sqrt{2}\cos\theta_p)/\sqrt{6},
\end{eqnarray}
$\theta_p$ is the singlet-octet mixing angle for pseudoscalar nonet, here we 
take $\theta_p=-15.5^\circ$\cite{angle}.
Introducing $r_1=g_2/g_1$, from Eqs. (14), (15) and (16), we have

\begin{eqnarray}
&&\frac{\Gamma(i\rightarrow\pi\pi)}{\Gamma(i\rightarrow K\overline{K})}=
3\left(
\frac{q_{i\pi\pi}}{q_{iK\overline{K}}}\right)^5
\frac{[X_i+(2X_i+\sqrt{2}Y_i)r_1]^2}
{[X_i+\sqrt{2}Y_i+(4X_i+2\sqrt{2}Y_i)r_1]^2},\\
&&\frac{\Gamma(i\rightarrow\eta\eta)}{\Gamma(i\rightarrow K\overline{K})}= 
2\left(\frac{q_{i\eta\eta}}
{q_{iK\overline{K}}}\right)^5
\frac{[\sqrt{2}\alpha^2X_i+2\beta^2Y_i+(\sqrt{2}X_i+Y_i)r_1]^2}
{[X_i+\sqrt{2}Y_i+(4X_i+2\sqrt{2}Y_i)r_1]^2},
\end{eqnarray}
where $i=f_2(1270),~f^\prime_2(1525)$, $q_{iP_1P_2}$ is the decay momentum 
for the decay mode $i\rightarrow P_1P_2$, 
\begin{eqnarray}
q_{iP_1P_2}=\sqrt{M^2_i-(M_{P_1}+M_{P_2})^2}\sqrt{M^2_i-(M_{P_1}-M_{P_2})^2},
\end{eqnarray}
$M_{P_1}$ and $M_{P_2}$ are the masses of the final pseudoscalar mesons
$P_1$ and $P_2$, respectively, and we take
$M_{K}=\sqrt{(M^2_{K^\pm}+M^2_{K^0})/2}$.

For the two-photon decays of $f_2(1270)$ and $f^\prime_2(1525)$, we 
have\cite{photon}

\begin{equation}
\frac{\Gamma(i\rightarrow\gamma\gamma)}
{\Gamma(a_2\rightarrow\gamma\gamma)}
=\frac{1}{9}
\left(\frac{M_i}{M_{a_2}}\right)^3
(5X_i+\sqrt{2}Y_i)^2,
\end{equation}
with $i=f_2(1270),~f^\prime_2(1525)$.

\section*{5. Comparison with the experimental data}

Now we wish to compared the theoretical results of Eqs. (18), (19) and
(21) with the available experimental data. From Eq. (9), the theoretical
results of Eq. (21) can be directly 
obtained as shown
in Table I. In order to obtain the theoretical results of Eqs. (18) and
(19), we should first determine the 
value 
of the unknown parameter $r_1$. In this procedure, we take 0.0092, the
experimental datum of
$\frac{\Gamma(f^\prime_2(1525)\rightarrow\pi\pi)}
{\Gamma(f^\prime_2(1525)\rightarrow K\overline{K})}$ as input, and 
determine the parameter $r_1$ to be the
value of $0.082$ or $0.161$ by solving the equation 
$\frac{\Gamma(f^\prime_2(1525)\rightarrow\pi\pi)}
{\Gamma(f^\prime_2(1525)\rightarrow K\overline{K})}=0.0092$. Using Eq. (9) 
and the value of $r_1$, the
theoretical results of Eqs. (18) and (19) are determined as shown in Table 
I.

From Table I, we find the theoretical results are in good agreement with
the experimental data, especially for $r_1=0.082$, i.e., the 
present experimental data support 
$|f_2(1270)\rangle=-0.991|N\rangle-0.135|S\rangle$ and
$|f^\prime_2(1525)\rangle=0.135|N\rangle-0.991|S\rangle$, which shows that 
the $f_2(1270)$ and
$f^\prime_2(1525)$ wave functions don't need other components such as 
glueballs or the $2~\pt$, $3~\pt$, ... $q\overline{q}$.

\section*{6. Summary and Conclusions}
Under the two-state mixing scheme, we determined the quarkonia content of
$f_2(1270)$ and $f^\prime_2(1525)$, and investigate the decays of the two 
states. The predicted results are in good
agreement with the experimental data. Our conclusions are as
follows:

1. The $f_2(1270)$ and $f^\prime_2(1525)$ wave functions don't need other 
components 
such as glueballs and the $2~\pt$, $3~\pt$, ... $q\overline{q}$.

2. $f_2(1270)$ is a nearly pure $1~\pt$ 
$(u\overline{u}+d\overline{d})/\sqrt{2}$
state ($\sim98.2\%$) and $f^\prime_2(1525)$ is a nearly pure $1~\pt$ 
$s\overline{s}$
state ($\sim98.2\%$). The isoscalar singlet-octet mixing angle for $1~\pt$
tensor nonet is determined to be the value of $27.5^\circ$.

\section*{6. Acknowledgments}
 
This work is supported by the National Natural Science Foundation of China
under Grant No. 19991487 and No. 19835060, and the foundation of Chinese
Academy of Sciences under Grant No. LWTZ-1298.

\tighten

\begin{table}
\begin{tabular}{cccccccc}
Decay&Exp.\cite{PDG}&\multicolumn{2}{c}{Pred.}&Decay&Exp.\cite{PDG}&
\multicolumn{2}{c}{Pred.}\\
\\
      &    &$r_1=0.082$ &$r_1=0.161$& & &$r_1=0.082$&$r_1=0.161$\\
\\
\hline
\\
$\frac{\Gamma(f_2\rightarrow\gamma\gamma)}
{\Gamma(a_2\rightarrow\gamma\gamma)}$&
$2.59\pm0.60$&2.67&2.67&
$\frac{\Gamma(f^\prime_2\rightarrow\gamma\gamma)} 
{\Gamma(a_2\rightarrow\gamma\gamma)}$&
$0.10\pm0.04$&0.09&0.09\\
\\
$\frac{\Gamma(f_2\rightarrow\pi\pi)}
{\Gamma(f_2\rightarrow K\overline{K})}$&
$18.41\pm2.52$&15.67&13.78&
$\frac{\Gamma(f^\prime_2\rightarrow\pi\pi)}
{\Gamma(f^\prime_2\rightarrow K\overline{K})}$&
$0.0092\pm0.0018$&0.0092$^\ast$&0.0092$^\ast$\\
\\
$\frac{\Gamma(f_2\rightarrow\eta\eta)}
{\Gamma(f_2\rightarrow K\overline{K})}$&
$0.10\pm0.03$&0.11&0.11&
$\frac{\Gamma(f^\prime_2\rightarrow\eta\eta)}
{\Gamma(f^\prime_2\rightarrow
K\overline{K})}$&  
$0.12\pm0.04$&0.10&0.11\\
\\
\end{tabular}
\vspace{1cm}
\caption{ The predicted and experimental results about the 
decays of $f_2(1270)$ and $f^\prime_2(1525)$. $f_2$ and $f^\prime_2$ 
respectively denote $f_2(1270)$ and $f^\prime_2(1525)$. ($^\ast$ input).}
\end{table}

\begin{figure}
\epsfysize=4.0 in
\vspace*{-1.3cm}
\centerline{\epsfig{figure=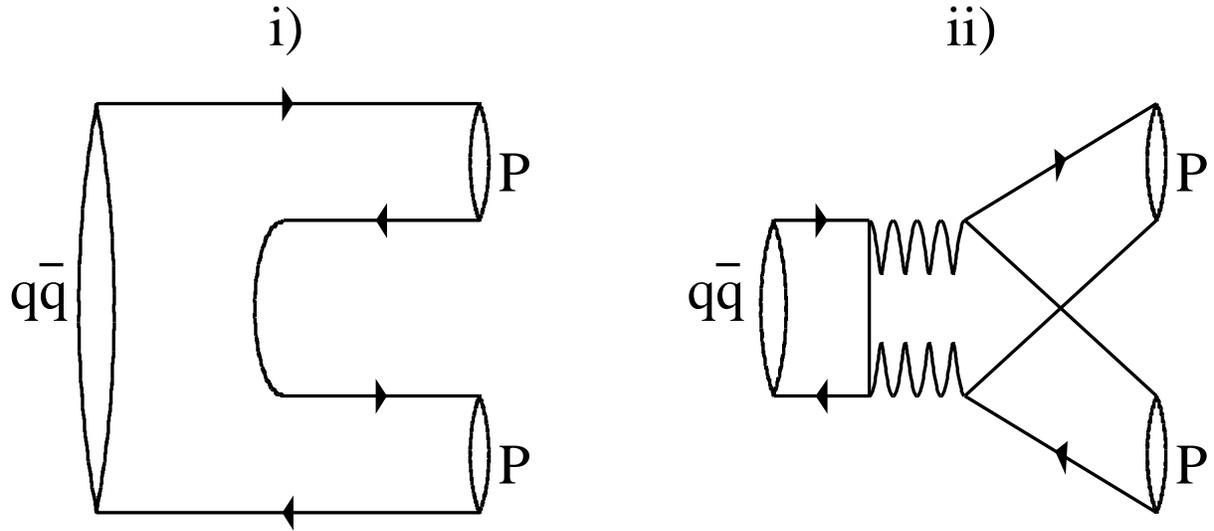}}
\caption{
The coupling modes of the quarkonia of $f_2(1270)$ and $f^\prime_2(1525)$ 
to the
pseudoscalar meson pairs ($PP$) considered in this work. i) The direct
coupling of the quarkonia components of the initial mesons to the final
pseudoscalar mesons occurring as the leading decay mechanism. ii) The
coupling of the quarkonia components of the initial mesons to the final
pseudoscalar mesons via intermediate gluons states occurring as the next
leading order decay mechanism.}
\end{figure}

\end{document}